\documentclass[11pt,epsf]{article}

\usepackage{epsfig}
\usepackage{amssymb}
\usepackage{graphicx}
\usepackage{color}
\usepackage{subfigure}

\begin{document}

\baselineskip 6mm
\renewcommand{\thefootnote}{\fnsymbol{footnote}}


\newcommand{\nc}{\newcommand}
\newcommand{\rnc}{\renewcommand}

\input amssym.def
\input amssym.tex


\newcommand{\tcb}{\textcolor{blue}}
\newcommand{\tcr}{\textcolor{red}}
\newcommand{\tcg}{\textcolor{green}}


\def\be{\begin{equation}}
\def\ee{\end{equation}}
\def\ba{\begin{array}}
\def\ea{\end{array}}
\def\bea{\begin{eqnarray}}
\def\eea{\end{eqnarray}}
\def\nn{\nonumber\\}


\def\ct{\cite}
\def\la{\label}
\def\eq#1{(\ref{#1})}


\def\a{\alpha}
\def\b{\beta}
\def\g{\gamma}
\def\G{\Gamma}
\def\d{\delta}
\def\D{\Delta}
\def\e{\epsilon}
\def\et{\eta}
\def\ph{\phi}
\def\Ph{\Phi}
\def\ps{\psi}
\def\Ps{\Psi}
\def\k{\kappa}
\def\l{\lambda}
\def\L{\Lambda}
\def\m{\mu}
\def\n{\nu}
\def\th{\theta}
\def\Th{\Theta}
\def\r{\rho}
\def\s{\sigma}
\def\S{\Sigma}
\def\ta{\tau}
\def\o{\omega}
\def\O{\Omega}
\def\pr{\prime}


\def\half{\frac{1}{2}}

\def\goto{\rightarrow}

\def\na{\nabla}
\def\grad{\nabla}
\def\curl{\nabla\times}
\def\div{\nabla\cdot}
\def\pa{\partial}

\def\bra{\left\langle}
\def\ket{\right\rangle}
\def\lb{\left[}
\def\lc{\left\{}
\def\ls{\left(}
\def\lp{\left.}
\def\rp{\right.}
\def\rb{\right]}
\def\rc{\right\}}
\def\rs{\right)}
\def\fr{\frac}

\def\vac#1{\mid #1 \rangle}


\def\td#1{\tilde{#1}}
\def\check{ \maltese {\bf Check!}}


\def\Tr{{\rm Tr}\,}
\def\det{{\rm det}}


\def\bc#1{\nnindent {\bf $\bullet$ #1} \\ }
\def\ch {$<Check!>$ }
\def\ss {\vspace{1.5cm}}

\begin{titlepage}

\hfill\parbox{5cm} { }

\vspace{25mm}

\begin{center}
{\Large \bf A Massive Quasi-normal Mode in the Holographic Lifshitz Theory}

\vskip 1. cm
  {Chanyong Park\footnote{e-mail : cyong21@ewha.ac.kr}}

\vskip 0.5cm

{\it Institute for the Early Universe, Ewha womans University, Daehyun 11-1, Seoul 120-750, Korea}
\\

\end{center}

\thispagestyle{empty}

\vskip2cm


\centerline{\bf ABSTRACT} \vskip 4mm

\vspace{1cm}
We investigate the holographic renormalization of the Einstein-Maxwell-dilaton
theory which provides an asymptotic Lifshitz geometry dual to a Lifshitz field theory.
In this case, the existence of a field combination with zero scaling dimension
causes an ambiguity in fixing local counter terms. Nevertheless, we show that
all possible local counter terms give rise to consistent thermodynamic quantities
with the Lifshitz black brane results.
In addition, we also study the retarded Green functions of the current and momentum
operator of a non-relativistic Lifshitz field theory.
In the non-zero momentum regime, the results show intriguingly that there exists a massive
quasi-normal mode whose effective mass is linearly proportional to temperature and that
even at zero temperature there exists a quasi-normal mode
in the non-relativistic Lifshitz medium.

\vspace{2cm}


\end{titlepage}

\renewcommand{\thefootnote}{\arabic{footnote}}
\setcounter{footnote}{0}


\section{Introduction}

After the Maldacena's conjecture \cite{Maldacena:1997re,Gubser:1998bc,Witten:1998qj}, there were a lot of works to understand strongly
interacting gauge theories via the dual classical gravity models. Recently, those works
were further generalized to the hyperscaling violation geometry in which the asymptotic
geometry is deviated from the AdS space and the boundary conformal symmetry is
broken \cite{Taylor:2008tg}-\cite{Kulkarni:2012in}. The study on the hyperscaling violation geometry is an interesting topic to
understand the gauge/gravity duality in depth and at the same time
to apply the holographic techniques to the real QCD \cite{Erlich:2005qh,Sakai:2004cn,Son:2002sd,Policastro:2002se,Policastro:2002tn,Kovtun:2003wp}, nuclear matter \cite{Lee:2009bya,Park:2009nb,Park:2011zp,Lee:2013oya} or condensed matter system \cite{Hartnoll:2008vx,Hartnoll:2009sz,Herzog:2009xv,McGreevy:2009xe,Horowitz:2010gk,Sachdev:2010ch}.

In the gauge/gravity duality point of view,
one of the interesting field contents on the gravity side is a nontrivial dilaton profile
because it is dual to a running gauge coupling of the dual gauge theory.
In addition, it would be possible to understand the nontrivial RG flow of the
strongly interacting systems through the holographic renormalization in the corresponding classical gravity theory
\cite{Verlinde:1999xm,Bianchi:2001kw,Skenderis:2002wp,Heemskerk:2010hk}.
There are several gravity models including a nontrivial dilaton field. One is a
relativistic non-conformal geometry which is the vacuum solution of the Einstein-dilaton
theory with a Liouville-type dilaton potential \cite{Taylor:2008tg}-\cite{Kulkarni:2012in}. 
In this case, a nontrivial dilaton profile
breaks the scaling symmetry of the metric without breaking the boundary Lorentz symmetry,
so the dual theory represents a relativistic non-conformal field theory \cite{Lee:2010qs,Lee:2010ii,Kulkarni:2012re,Park:2012cu,Kulkarni:2012in}.
The other interesting example is the Lifshitz geometry which was invented to understand the
non-relativisitic features by using the holographic techniques. There were several
gravity models leading to the asymptotic Lifshitz geometry, for examples, the gravity
theory with various higher form fields \cite{Kachru:2008yh} or  with a massive vector field 
\cite{Balasubramanian:2009rx,Korovin:2013nha}.
In this paper, we will concentrate on another model, the so-called Einstein-Maxwell-dilaton theory, in which
a bulk gauge field as well as a nontrivial dilaton profile are introduced \cite{Pang:2009wa,Park:2013goa}.

In an asymptotic AdS background without a dilaton field, a time component of a bulk gauge
field is dual to the number operator of matter in the dual field theory and its geometry
can be described by the thermal charged AdS for the confining phase at low temperature
or the Reinssner-Nordstr\"{o}m black brane for the deconfining phase at high temperature \cite{Lee:2009bya,Park:2009nb,Park:2011zp,Lee:2013oya}.
In an asymptotic Lifshitz geometry, although there exists a bulk gauge field,
it does not provide a new hair to the black brane solution. Instead, it together with the nontrivial
dilaton field  changes the asymptotic geometry to the Lifshitz one.
Furthermore, the anisotropic scaling symmetry of the Lifshitz geometry appears due to the
breaking of the boost symmetry between time and spatial coordinates caused by the background gauge field.
As a result, a general solution of the Einstein-Maxwell-dilaton theory is
given by a Schwarzschild-type black brane with the Lifshitz-type scaling symmetry.
Its dual field theory can be reinterpreted as a gauge theory
with Lifshitz matter whose dispersion relation is governed by the Lifshitz-type field theory.
In this case, since the boundary value of the gauge field is not a free parameter, the density
(or the chemical potential) of  Lifshitz matter is fixed by the intrinsic parameters of the
Einstein-Maxwell-dilaton theory which may be reinterpreted as a microcanonical ensemble.

In order to understand the thermodynamic properties and their RG flow, it is interesting
to study the holographic renormalization. The holographic renormalizations of other models
describing the asymptotic Lifshitz geometry have been already investigated \cite{Balasubramanian:2009rx,Korovin:2013nha,Ross:2009ar,Korovin:2013bua}. 
Here, we concentrate on
the holographic renormalization of the Einstein-Maxwell-dilaton theory. In this model,
the local counter terms are not uniquely fixed due to the
existence of a combination whose leading term at the asymptotic boundary
has a zero scaling dimension. When we consider several lowest order counter terms,
we find that all possible combinations in the local counter terms give rise
to the same thermodynamics consistent with the Lifshitz black brane result.
If we further regard the RG flow, the fact that the local counter term
is not unique implies that the RG flow can not be also determined uniquely. Unfortunately,
we do not have a clear idea yet how to resolve this problem. It may require more deep understanding about the microscopic aspects of the dual Lifshitz theory.
We leave it as a future work.

In the zero momentum limit of the dual Lifshitz theory, the electric properties 
carried by Lifshitz matter and impurity were studied \cite{Pang:2009wa,Park:2013goa}. 
Here, we further investigate the transport coefficients carried by Lifshitz matter in the non-zero momentum limit
of the non-relativistic Lifshitz theory ($z=2$).
Due to the existence of the background gauge field,
a charge current is usually mixed with a momentum operator. We explicitly calculate
the retarded Green functions of a charge current and momentum operator in the hydrodynamic limit.
The results show that the shear viscosity is linearly proportional to temperature and
saturates the lower bound of $\et/s$ \cite{Pang:2009wa}. Interestingly,
the retarded Green functions of the non-relativistic Lifshitz theory show the existence of a
massive quasi-normal mode which has a thermal effective mass linearly proportional to temperature and a constant momentum diffusion constant.

The rest of paper is organized as follows. In Sec. 2, we rederive the thermodynamics of the Lifshitz
black brane by using the holographic renormalization. Although the local counter terms are
not determined uniquely, we show that all possible counter terms lead to the consistent
thermodynamics with the black brane result.
On this background, we further investigate the transport coefficients carried by a Lifshitz
matter in Sec. 3. Intriguingly, the current-current and
momentum-momentum retarded Green functions show a massive quasi-normal mode whose
mass is linearly proportional to temperature.
In Sec. 4, we finish this work with some concluding remarks.


\section{A holographic renormalization of a Lifshitz theory}

Let us start with briefly summarizing the Lifshitz black brane geometry, its thermodynamics and
our notations with a Lorentzian signature.
The action for an Einstein-Maxwell-dilaton theory with a negative cosmological constant $\L$
is given by \cite{Taylor:2008tg,Pang:2009wa,Park:2013goa}
\be			\la{act:original}
S_{EMd} = \frac{1}{16 \pi G} \int d^{4} x \sqrt{-g} \ls  R - 2 \L
- \half \pa_{\m} \ph \ \pa^{\m} \ph - \frac{1}{4} e^{\l \ph} F_{\m\n} F^{\m\n} \rs ,
\ee
which is believed to be the dual gravity of a Lifshitz-type field theory.
The Einstein and Maxwell equation are given by 
\bea
R_{\mu\nu}-\frac{1}{2}Rg_{\mu\nu}+ g_{\mu\nu} \L &=& \half \partial_{\mu}\phi
\partial_{\nu}\phi - \frac{1}{4}  g_{\mu\nu}(\pa \phi)^{2}+ \half  e^{\l \phi}
F_{\mu\lambda}{F_{\nu}}^{\lambda}
-\frac{1}{8} g_{\mu\nu}e^{\l \phi} F^{2},   \la{eq:Einstein} \\
\partial_{\mu}(\sqrt{-g}\partial^{\mu}\phi) &=& \frac{\l}{4}\sqrt{-g}e^{\l \phi}F^{2} ,
\la{eq:Maxwell}
\eea
and the equation of motion for dilaton is
\be
0 = \partial_{\mu}(\sqrt{-g}e^{\l \phi}  F^{\m \n})  . \la{eq:dilaton}
\ee
These equations allow a Schwarzschild-type black brane solution
\bea			\la{sol:lifshitz}
ds^2 &=& - r^{2z} f(r) dt^2 + \frac{dr^2}{r^2 f(r)} + r^2 (dx^2 + dy^2 )  , \nn
\ph (r) &=& - \fr{4}{\l} \log r ,\nn
F_{rt} &=& \pa_r A_t = q \ r^{z+1} ,
\eea
with
\bea
f(r) &=& 1 - \frac{r_h^{z+2}}{r^{z+2}} , \nn
\l &=& \frac{2}{\sqrt{z-1}} , \nn
q &=& \sqrt{2 (z-1) (z+2)} ,\nn
\L &=& - \frac{(z+1)(z+2)}{2} ,
\eea
where $r_h$ denotes an event horizon and $z$ is a dynamical exponent.
We simply call it the Lifshitz black brane. Its Hawking temperature
from the surface gravity is
\be			\la{res:hawkingtemp}
T_H = \frac{z+2}{4 \pi} r_h^{z} ,
\ee
and the Bekenstein-Hawking entropy becomes
\be				\la{res:thentropy}
S_{BH} = \frac{V_2}{4 G} r_h^2 ,
\ee
where $V_2$ means a spatial volume of the boundary space.

The Lifshitz black brane thermodynamics can be reinterpreted as that of the dual
Lifshitz field theory following the gauge/gravity duality.
From the thermodynamic law together with information of the Hawking temperature and
Benkenstein-Hawking entropy, one can easily read other thermodynamic quantities.
The internal energy $E$ and the free energy $F$ are
\bea
E &=& \fr{V_2}{8 \pi G} r_h^{z+2} , \nn
F &=& - \fr{z V_2}{16 \pi G} r_h^{z+2} .
\eea
In addition, the pressure is given by $P = - \pa F/ \pa V_2$ which satisfies the Gibbs-Duhem
relation, $E + P V_{2} = T_H S_{BH}$.

The above thermodynamic results of the dual Lifshitz theory are just reinterpretation
of the Lifshitz black brane geometry.
Using the holographic renormalization method \cite{Park:2012cu,Balasubramanian:1999re,Son:2006em,Batrachenko:2004fd}, it was shown in a non-conformal geometry as well as an asymptotic AdS geometry that the boundary stress tensor 
gives rise to the same thermodynamics as the black brane solution.
Furthermore, as was shown in the different models for the Lifshitz geometry 
\cite{Balasubramanian:2009rx,Ross:2009ar}, it is also true in the Einstein-Maxwell-dilaton gravity theory.
These facts indicate that the assumption of the gauge/gravity duality is self-consistent 
even in the asymptotic Lifshitz geometry.
In this paper,  following the gauge/gravity duality assumption we will show that the holographic 
renormalization of the Einstein-Maxwell-dilaton gravity can reproduce the same thermodynamics 
of the Lifshitz black brane.
To do so, we should first assume that similar to the AdS case the on-shell gravity action can be 
mapped to a boundary term proportional to the free energy of the dual field theory even in the non-AdS space. 
In this procedure, an appropriate holographic renormalization is required to remove UV divergences.
After the holographic renormalization, we will show that the boundary stress tensor leads to the
consistent thermodynamics in spite of an ambiguity in choosing the local counter terms. 

From now on, we use the Euclidean signature \cite{Park:2012cu} because it is more convenient
for later comparison.
The Euclidean action can be easily obtained
by applying the Wick rotation $t \to - i \ta$, in which the time component of the gauge field
should be also rotated like $A_t \to i A_{\ta}$.
Then, the Euclidean Einstein-Maxwell-dilaton action can be rewritten as
\be			\la{act:EMd}
S_{EMd} = - \frac{1}{16 \pi G} \int_{\cal M} d^{4} x \sqrt{g} \ls  R - 2 \L
- \half \pa_{\m} \ph \ \pa^{\m} \ph - \frac{1}{4} e^{\l \ph} F_{\m\n} F^{\m\n} \rs ,
\ee
where $g_{\m \n}$ is the Euclidean metric. Since the Euclidean action has the same as \eq{act:original}
up to an overall minus sign, all equations of motion are also the same as the previous ones in
\eq{eq:Einstein}, \eq{eq:Maxwell} and \eq{eq:dilaton}.
By explicit calculation, one can easily check that the Einstein and Maxwell equations
are really satisfied by the Euclidean metric
\be		\la{sol:Emetric}
ds^2 = r^{2z} f(r) \ d {\ta}^2 + \frac{dr^2}{r^2 f(r)} + r^2 (dx^2 + dy^2 ) ,
\ee
and the Wick-rotated time component gauge field
\be         \la{sol:Emaxwell}
A_{\ta} = -\fr{i \ q}{z+2}\ls  r^{z+2}  -  r_h^{z+2} \rs ,
\ee
where the last constant term is introduced for a well-defined norm of $A_{\ta}$ at the
black brane horizon.

For a well-defined metric variation of the action,
the Gibbons-Hawking term must be added
\be
S_{GH} = \fr{1}{8 \pi G} \int_{\pa {\cal M}} d^{3} x \sqrt{\g} \ \Th ,
\ee
where $\Th$ is an extrinsic curvature scalar and $\g_{ab}$ is an induced metric
on the boundary. An extrinsic curvature tensor $\Th_{\m\n}$ is defined by
\be
\Th_{\m\n} = - \half \ls \na_{\m} n_{\n} + \na_{\n} n_{\m} \rs ,
\ee
where $\na_{\m}$ and $n_{\n}$ denote a covariant derivative and unit
normal vector respectively. Since the Gibbons-Hawking term is a boundary term,
it does not affect on the equations of motion. Furthermore, one can also add an additional boundary term
called the Neumannizing term \cite{Balasubramanian:2009rx,Adams:2008wt}
\be
S_{N} = - \fr{c_N}{16 \pi G} \int_{\pa {\cal M}} d^3 x \sqrt{\g} \ e^{\l \ph} \ n^{\m} A^{\n} F_{\m \n} ,
\ee
which determines the boundary condition of the bulk gauge field. For $c_N = 0$,
the gauge field satisfies the Dirichlet boundary condition, whereas $c_N = 1$ when
the Neumann boundary condition is imposed. Here, we choose a Dirichlet boundary condition,
so our starting action for the Lifshitz theory is
\be			\la{act:starting}
S = S_{EMd} + S_{GH} .
\ee
Although this action has a well-defined metric variation, it still suffers from the UV divergence
at $r_0 \to \infty$ where $r_0$ denotes the UV cutoff or the position of boundary. In addition,
the boundary terms caused by the variations with respect to other matter fields
can also yield the UV divergences.
These divergences can be removed by the holographic renormalization with appropriate local counter
terms. What are the correct local counter terms?
Following the gauge/gravity duality, the renormalized on-shell gravity action is proportional to the free energy of the dual field theory and the boundary energy-momentum tensor derived
from it can be identified with that of the dual field theory. Therefore, the renormalized
action and its boundary energy-momentum tensor should be independent of the UV cutoff
introduced by hand and finite even at $r_0 \to \infty$.

The on-shell action without local counter terms at the UV boundary,
after inserting solutions to \eq{act:starting}, leads to
\be
S_{on} =  \b V_2 \ls - \fr{(1 + z)}{8 \pi G} r_0^{z+2}   + \fr{  z }{16 \pi G } r_h^{z+2} \rs
+ {\cal O} \ls \fr{1}{r_0^{z+2}} \rs ,
\ee
where $\b$ is the Euclidean time periodicity proportional to the inverse temperature
and $V_2$ is a regularized spatial volume of the boundary space.
The variation of the on-shell action with respect to the boundary metric gives rise to the
boundary energy-momentum tensor
\be
{{T}^a}_b = - \fr{1}{8 \pi G} \int d^{2} x \sqrt{\g} \ \g^{ac} \ls \Th_{cb}
-  \g_{cb} \Th \rs ,
\ee
where $a$ and $b$ means the boundary coordinates. Note that there is no contribution from the kinetic terms of the dilaton and vector field.
The explicit energy and pressure of the dual Lifshitz theory read
\bea  \la{res:bbthermodyn}
E &=& {T^{\ta}}_{\ta}  =   -  \fr{V_2 }{4 \pi G} r_0^{z+2}  +\fr{V_2 }{4 \pi G}  r_h^{z+2}
+ {\cal O} \ls \fr{1}{r_0^{z+2}}  \rs , \nn
P_i &=& - \fr{{T^{i}}_{i}}{V_{2}}  = \fr{(1 + z)}{8 \pi G} r_0^{z+2}  - \fr{  z }{16 \pi G } r_h^{z+2}  + {\cal O} \ls \fr{1}{r_0^{z+2}} \rs ,
\eea
where an index $i$ implies the spatial direction.
It is worth to note that the free energy given by $F = \fr{S_{on}}{\b}$ 
is exactly the same as $- P_i V_2$. The above results are unrenormalized ones so that
they suffer from the UV divergences at $r_0 \to \infty$, as mentioned before. 
Since the pressure is proportional to the free energy,
at least two
local counter terms are required to make the energy-momentum tensor finite at the UV cutoff.

Similar to the energy-momentum tensor, the variations with respect to the other matter fields
also suffer from the UV divergences.
Varying the unrenormalized on-shell action with respect to the matter fields gives rise to
\bea
\fr{\pa S_{on}}{\pa \ph} &=& -\frac{ \sqrt{z-1} \beta V_2}{8 \pi  G} \ r_0^{z+2}
+ \frac{ \sqrt{z-1} \beta V_2 }{8 \pi  G} \ r_h^{z+2}  + \cdots, \nn
\fr{\pa S_{on}}{\pa A_{\ta}} &=& -\frac{i \sqrt{z-1} \sqrt{z+2} \beta V_2  }{8 \sqrt{2} \pi  G}
+ \cdots ,
\eea
where the ellipses imply the higher order corrections which vanish for $r_0 \to \infty$.
For the well-defined variations and finiteness, these terms should be also removed by three more counter terms.
As a result, the renormalization of the on-shell action requires five counter terms if all
constraints for eliminating the UV divergences are independent.
Actually, only three counter terms, as will be shown, are sufficient for the renormalization
due to the redundancies of the constraints.
Then, what kinds of the local counter term are possible? 

In order to remove the UV divergences,
the counter terms having the same leading divergence should be taken into account.
There exist infinitely many possible counter terms generating the same divergence
\be
S_{possible} = \sum_{i=0}^{\infty} \fr{c_i }{8 \pi G} \int d^3 x \sqrt{\g}    \ls e^{\l \ph} A^2 \rs^i ,
\ee
where $\sqrt{\g} \sim r_0^{z+2}$ and the leading term of $e^{\l \ph} A^2$ is given by
a constant. Let us first assume that all constraints are independent, then
the renormalized action contains at most five counter terms as previously mentioned.
Since infinite many counter terms are possible, there exists an ambiguity in choosing five counter terms. Here, we simply choose the five lowest order counter terms
\be        \la{act:counter}
S_{ct} = \sum_{i=0}^{4} \fr{c_i }{8 \pi G} \int d^3 x \sqrt{\g}    \ls e^{\l \ph} A^2 \rs^i .
\ee
where $c_i$ are coefficients to be determined. If adding more counter terms, the coefficients of
them are not uniquely fixed because of the lack of constraints.
Similarly, if the previous constraints are not independent, all coefficients in \eq{act:counter} can not be
fixed exactly.  This fact also generates the same ambiguity
in choosing appropriate counter terms.
As will be shown subsequently, only three constraints in the Lifshitz geometry are independent
so that two coefficients can not be determined fully.
In order to get rid of these two redundancies, one can simply set two of them in \eq{act:counter}
to be zero.
It should be noted that $c_0$ must have a non-vanishing value because the on-shell action of the AdS space ($z=1$)
is renormalized only by $c_0$. As a result, due to the non-vanishing $c_0$
there are six possibilities in choosing three counter terms.

Here, to show the redundancies explicitly let us start with five counter terms.
The resulting renormalized action with an Euclidean signature is described by
\be
S_{ren} = S_{EMd} + S_{GH} + S_{ct} ,
\ee
where the counter terms is given by \eq{act:counter}.
After substituting the solutions, \eq{sol:Emetric} and \eq{sol:Emaxwell}, into the renormalized
action and expanding it near the UV cutoff, the following five constraint equations
are derived
\bea
0
&=& -2 z^4-16 z^3-48 z^2-64 z-32 + (z+2)^4 c_0 +2  (z+2)^3 (z-1) c_1
-12  \left(z^2+z-2\right)^2 c_2  \nn
&& +40  (z+2) (z-1)^3 c_3 -112  (z-1)^4  c_4    ,  \la{const:first}    \\
0
&=& -z^5-9 z^4-32 z^3-56 z^2-48 z-16 + (z+2)^4 c_0 - 2  (z-1) (z+2)^3 c_1
   + 4  \left(z^2+z-2\right)^2 c_2 \nn
&& -8  (z-1)^3 (z+2) c_3 + 16 (z-1)^4  c_4, \la{const:second} \\
0
 &=& -z^4-8 z^3-24 z^2-32 z-16 -4  (z+2)^3 c_1 +16  (z-1) (z+2)^2 c_2 -48
 \left(z^3-3 z+2\right) c_3  \nn
 && + 128  (z-1)^3 c_4, \la{const:third} \\
0
 &=& z^4+8 z^3+24 z^2+32 z + 16 +6 (z+2)^3  c_1 -40  (z-1) (z+2)^2 c_2 + 168 \left(z^3-3 z+2\right)  c_3 \nn
 && -576  (z-1)^3 c_4 , \la{const:fourth} \\
0
 &=& -z^4-8 z^3-24 z^2-32 z-16 -4  (z+2)^3 c_1 +16 (z-1) (z+2)^2 c_2  -48
  \left(z^3-3 z+2\right) c_3 \nn
  && +128  (z-1)^3 c_4 . \la{const:fifth}
\eea
In the above, the first two equations describe vanishing of the divergence in energy
and pressure. The third and fourth equations come from the well-defined variation
with respect to the dilaton field. The remaining is the condition
for the $A_{\ta}$ variation. Since the third constraint in \eq{const:third} is the same as the fifth
constraint in \eq{const:fifth}, all constraints are not independent. Furthermore, there exist another redundancy because the above constraints
automatically satisfy the following relation
\be
0 = \eq{const:first} - \eq{const:second} + (z-1) \ \eq{const:third} .
\ee
Therefore, as mentioned before, only three of them are independent. This means that three
counter terms are sufficient in the holographic renormalization of the Lifshitz theory.

Which coefficients can we set to be zero? This is an important question to fix the counter
terms uniquely. However, at the present stage unfortunately we have no concrete idea
for choosing three of them.
In this paper, instead of resolving the uniqueness problem of the counter terms, we will investigate
whether the physical properties of the dual theory crucially depend on the choice of the
counter terms or not.  In what follows, we summarize six different parameter solutions, which
are only allowed cases when one starts with five lowest counter terms \\

1) For $c_3 =c_4 =0$,
\be			\la{res:c3c40}
c_0 = \fr{13 + 3 z}{8} \ , \quad  c_1 = - \fr{3 (2 + z)}{8} \ ,
\quad c_2 = - \fr{(z+2)^2}{32 (z-1 )}  , \quad
c_3 = 0 \ , \ {\rm and} \quad  c_4 = 0  .
\ee

2) For $c_2 =c_4 =0$,
\be
c_0 = \frac{1}{12} (5 z+19)\ , \quad
c_1 =   -\frac{5}{16} (z+2) \ , \quad
c_2 = 0 , \quad
c_3 =  \frac{(z+2)^3}{192 (z-1)^2} \ , \ {\rm and} \quad
c_4 = 0  .
\ee

3) For $c_1 =c_4 =0$,
\be
c_0 = \frac{1}{8} (5 z+11)    \ , \quad
c_1 =  0 \ , \quad
c_2 =   \frac{5 (z+2)^2}{32 (z-1)}      , \quad
c_3 =    \frac{(z+2)^3}{32 (z-1)^2}        \ , \ {\rm and} \quad
c_4 = 0  .
\ee

4) For $c_2 =c_3 =0$,
\be
c_0 =    \frac{1}{16} (7 z+25)           \ , \quad
c_1 =     -\frac{7}{24} (z+2)           \ , \quad
c_2 = 0  , \quad
c_3 = 0 \ , \ {\rm and} \quad
c_4 =  -\frac{(z+2)^4}{768 (z-1)^3}  .
\ee

5) For $c_1 =c_3 =0$,
\be
c_0 =   \frac{1}{32} (21 z+43)                   \ , \quad
c_1 = 0 \ , \quad
c_2 =    \frac{7 (z+2)^2}{64 (z-1)}                    , \quad
c_3 = 0 \ , \ {\rm and} \quad
c_4 =  -\frac{3 (z+2)^4}{512 (z-1)^3} .
\ee

6) For $c_1 =c_2 =0$,
\be   		\la{res:c1c20}
c_0 =   \frac{1}{48} (35 z+61)                 \ , \quad
c_1 = 0 \ , \quad
c_2 = 0 , \quad
c_3 =     -\frac{7 (z+2)^3}{96 (z-1)^2}               \ , \ {\rm and} \quad
c_4 =     -\frac{5 (z+2)^4}{256 (z-1)^3}  .
\ee \\

Note that for $z=1$, $A_{\ta}$ automatically vanishes so that only $c_0$ is
required for the renormalization. In this case, $c_0$ reduces to $2$ which is the case
for the $AdS_4$ space \cite{Balasubramanian:1999re}.
In general cases, the internal energy and pressure depending on five coefficients read
from the boundary energy-momentum tensor
\bea
E &=& {T^{\ta}}_{\ta} \nn
&=&\frac{(z+2) \left[ (z+2) \lc 4 (z+2)^2 - (z+2)^2 c_0 -6  \left(z^2+z-2\right) c_1 +60  (z-1)^2 c_2
 \rc -280  (z-1)^3 c_3 \right]  }{16 \pi  G (z+2)^4}  \ r_h^{z+2}\nn
   && + \fr{1008  (z-1)^4 c_4 }{16 \pi  G (z+2)^4}  \ r_h^{z+2} , \nn
P_i &=& - \fr{{T^{i}}_{i}}{V_{2}}  \nn
&=& \frac{ (z+2) \lb (z+2) \lc - z (z+2)^2 + (z+2)^2 c_0  -6  \left(z^2+z-2\right) c_1 +20  (z-1)^2 c_2  \rc -56  (z-1)^3 c_3 \rb  }{16 \pi  G (z+2)^4} \ r_h^{z+2} \nn
&&  +\fr{ 144
   (z-1)^4 c_4 }{16 \pi  G (z+2)^4} \  r_h^{z+2} .
\eea
Using the above results given in \eq{res:c3c40}$\sim$\eq{res:c1c20},
the resulting thermodynamic quantities, the free energy and energy-momentum tensor
at $r_0 = \infty$, lead to the same result in all cases
\bea
F &=& - \fr{z V_2}{16 \pi G} r_h^{z+2} , \nn
E &=& \fr{V_2}{8 \pi G}  r_h^{z+2}, \nn
P_i &=& \fr{z }{16 \pi G} r_h^{z+2} ,
\eea
which are perfectly matched to the black brane thermodynamics in \eq{res:bbthermodyn}.
These results show that although the local counter terms are not
fixed uniquely, the holographic renormalization of the Lifshitz black brane
leads to the correct boundary energy-momentum tensor of the dual Lifshitz theory.
Furthermore, the holographic renormalization shows the self-consistency of the gauge/gravity duality 
even in the asymptotic Lifshitz geometry.

\section{A massive quasi-normal mode in the non-relativistic theory}

In the hydrodynamics of the relativistic quantum field theory,
it was well-known that a momentum
diffusion constant can be represented by the background thermodynamic quantities 
\be			\la{res:QFT}
{\cal D} = \fr{\et}{\e + P} ,
\ee
where $\et$, $\e$ and $P$ are the shear viscosity, energy density and pressure.
This result has been checked in the dual conformal field theory of the asymptotic AdS space
by using the holographic hydrodynamics \cite{Hartnoll:2009sz,Herzog:2009xv,McGreevy:2009xe,Horowitz:2010gk,Sachdev:2010ch}. Furthermore, it was also shown that
this relation is true even in the relativistic non-conformal theory dual to a non-AdS
geometry \cite{Park:2012cu}.  However, we can not expect that the above momentum
diffusion constant is still valid in the asymptotic Lifshitz geometry because of the
breaking of the Lorentz symmetry.
Therefore, it is interesting to calculate the momentum diffusion constant of the dual Lifshitz
theory. In this section, we will investigate
the holographic hydrodynamics of the Lifshitz geometry, especially the non-relativistic case
($z=2$).

In general, if there exists a nonzero background gauge field,
the shear mode of the metric fluctuation, $g^x_t$ and $g^x_y$, and the transverse mode of the gauge field
fluctuation $a_x$ are usually coupled to each other.  In order to evaluate the
momentum diffusion constant of the non-relativistic Lifshitz theory, one should turn on
the gauge and metric fluctuations simultaneously.
Now, let us expand all fluctuations as Fourier modes
\bea
a_x (t,y,r) &=& \int \fr{d \o \ d k}{(2 \pi)^2} \ e^{- i \o t + i k y} \ a_x (\o,k,r) , \nn
g^x_t (t,y,r) &=& \int \fr{d \o \ d k}{(2 \pi)^2} \ e^{- i \o t + i k y} \ g^x_t (\o,k,r) ,\nn
g^x_y (t,y,r) &=& \int \fr{d \o \ d k}{(2 \pi)^2} \ e^{- i \o t + i k y} \
 g^x_y (\o,k,r)   ,
\eea
For a general $z$, shear modes are governed by
\bea
0 &=&  \frac{\o}{r^{2 z-2}} {g^x_t}^{\pr}  + k f {g^x_y}^{\pr} +  \frac{q \o}{r^{z +3}}  a_x  ,
\la{eq:constraint} \\
0 &=& {g^x_t}^{\pr\pr} + \frac{(5-z)  }{r} {g^x_t}^{\pr}
-\frac{  k^2 }{ r^4 f } g^x_t  -\frac{k \o }{r^4 f } g^x_y + \frac{q}{r^{5-z}} a_x^{\pr} ,
\la{eq:gtx} \\
0 &=& {g^x_y}^{\pr\pr} + \frac{ r f^{\pr} +(z+3) f  }{r f } {g^x_y}^{\pr}
+ \frac{ \o^2 }{ r^{2z +2}  f^2} {g^x_y}   +\frac{k \o  }{r^{2 z+2} f ^2} {g^x_t} ,
\la{eq:gxy}
\eea
and the Maxwell equation for $a_x$ is given by
\be	 \la{eq:a}
0= a_x^{\pr \pr} +\frac{ r f^{\pr} + (z-3) f  }{r  f} a_x^{\pr}
+\frac{r^2 \o^2-k^2 r^{2 z} f  }{r^{2 z + 4} f ^2} a_x +\frac{q r^{3-z} }{f } {g^x_t}^{\pr} .
\ee
Note that for $z=1$, since $q=0$, the transverse mode is decoupled from
shear modes. 
In equations of the shear modes, only two of them are independent because combining
\eq{eq:constraint} and \eq{eq:gtx} leads to the rest one \eq{eq:gxy}.
From now on we concentrate on the $z=2$ case because its dual theory is described by a
non-relativistic quantum field theory. Combining \eq{eq:constraint}, \eq{eq:gtx} and \eq{eq:a}
leads to
\bea			\la{eq:solve1}
0 &=&\Phi'' +\left(\frac{f'}{f}+\frac{7}{r}\right) \Phi ' +   \left(
\frac{\omega ^2}{r^6 f^2} + \frac{3 f'}{r f}-\frac{k^2}{r^4 f}
- \fr{q^2}{r^2 f} +\frac{9}{r^2}\right) \Phi  \nn
&& +  \fr{q}{r^4}  \ls 2a_x' + \frac{k^2 }{r^3
   f}  a_x \rs ,
\eea
where $\Ph = {g^x_t} '$, and \eq{eq:a} reduces to
\be		\la{eq:solve2}
0 = a_x'' +  \left(\frac{f'}{f}-\frac{1}{r}\right) a_x '  +
   \left(\frac{\omega ^2}{r^6 f^2}-\frac{k^2}{r^4 f}\right) a_x +\frac{q r  }{f} \Phi .
\ee

Near the event horizon, the leading solutions satisfying the incoming boundary condition
read
\bea			\la{sol:horizon}
a_x (r_h) &=& F_0 \  f^{-  i \fr{\o}{4 r_h^2} } , \nn
\Ph (r_h) &=& G_0 \ f^{-  i \fr{\o}{4 r_h^2} } ,
\eea
where $F_0$ and $G_0$ are two integration constants. Furthermore, in the hydrodynamic
limit ($\o \sim k^2 \ll T_H$),
the perturbative solutions near the horizon can be expanded into
\bea		\la{sol:nearhorizonsol1}
a_x (r_h) &=& f^{-  i \fr{\o}{4 r_h^2} } \ls F_0 \ \d (k) + \o \ F_1 (r) \ \d (k) +  k^2 \ F_2 (r) + \cdots \rs , \nn
\Ph (r_h) &=& f^{-  i \fr{\o}{4 r_h^2} } \ls G_0 \ \d (k)  + \o \ G_1 (r) \ \d (k) + k^2 \ G_2 (r) + \cdots \rs ,
\eea
where $\d (k)$ implies the zero momentum mode. The zero momentum modes do not coupled to 
$g^x_y$.
Solving \eq{eq:solve1} and \eq{eq:solve2} perturbatively, the solutions $F_0$ and $G_0$, which
are regular at the horizon,  are given by
\bea
F_0 &=& \left( r^4 + r_h^4 \right)  c_1  , \nn
G_0 &=& -\frac{2 \sqrt{2} \left( r^4 + r_h^4 \right)}{r^3}  c_1 ,
\eea
where $c_1$ is an undetermined integration constant.
In order to satisfy \eq{sol:horizon} at the horizon, higher order solutions should be
vanishing as well as regular. The solutions satisfying these constraints lead to
\bea
F_1 (r) &=& -\frac{i c_1}{4 \ r_h^2}   \ \lb 2  \left(r^4+ r_h^4\right) \ls 2 \log r + \log 2 \rs
- (r^2 - r_h^2)^2
-2 \left(r^4+r_h^4\right) \log \left(r^2+r_h^2\right)  \rb , \nn
G_1 (r) &=& \frac{i c_1}{\sqrt{2} \ r_h^2  r^3} \ \lb 2 \left(r^4+r_h^4\right) \ls 2 \log r + \log 2 \rs
- (r^2 - r_h^2)^2
-2 \left(r^4+r_h^4\right) \log \left(r^2+r_h^2\right)  \rb  , \nn
F_2 (r) &=& -\frac{c_1}{32 r_h^2} \lb
\pi^2 \ls r^4 +r_h^4 \rs + 2 \ls r^4 + 3 r_h^4 \rs \ls r^4 - r_h^4\rs
+ 2 \left(r^4 +r_h^4 \right) \lc {\rm Li}_2\left(\frac{r^4}{r_h^4}\right)-4 {\rm Li}_2\left( \frac{r^2}{r_h^2}\right) \rc  \rp  \nn
&& \qquad  \qquad  \lp  -
8  \log \ls \fr{r}{r_h} \rs \lc  \left(r^4 + r_h^4 \right)  \log \ls r^4-r_h^4 \right)
+   2  r^2 r_h^2 \rc  \rb , \nn
G_2 (r) &=& \frac{c_1}{8 \sqrt{2} \ r_h^2 r^3 }
\lb
\pi^2 \ls r^4 +r_h^4 \rs + 2  \ls r^2 - r_h^2 \rs^2
+ 2 \left(r^4 +r_h^4 \right) \lc {\rm Li}_2\left(\frac{r^4}{r_h^4}\right)-4 {\rm Li}_2\left( \frac{r^2}{r_h^2}\right) \rc  \rp  \nn
&& \qquad  \qquad \quad \lp  -
8  \log \ls \fr{r}{r_h} \rs \lc  \left(r^4 + r_h^4 \right)  \log \ls r^4-r_h^4 \right)
+   2  r^2 r_h^2 \rc  \rb  ,
\eea
where ${\rm Li}$ denotes a polylogarithm function. The zero momentum modes
$F_0 (r)$ and $F_1 (r)$ coincide with the results obtained in the zero momentum limit \cite{Park:2013goa}.

Near the horizon, the remaining $g^x_y$ can be again written as sum of the zero and nonzero
momentum mode
\be			\la{sol:nearhorizonsol2}
g^x_y = f^{-  i \fr{\o}{4 r_h^2} } \ls   H_0 \ \d (k)  + \o \ H_1 (r) \ \d (k) + \ H_2 (\o,k,r) + \cdots \rs .
\ee
The zero momentum modes are governed by
\be			\la{eq:zeromom}
0 = \ps^{\pr\pr} + \ls \frac{f^{\pr}}{f} + \fr{5}{r} \rs \ps^{\pr}
+ \frac{ \o^2 }{ r^6  f^2} \ps   .
\ee
where $\ps (\o,r) $ denotes the zero momentum modes, $\ps = f^{-  i \fr{\o}{4 r_h^2} }
\ls H_0 + \o \ H_1 (r) + {\cal O} (\o^2) \rs $. This zero momentum mode 
is decoupled from others and determines the shear viscosity of the dual system.
The regularity and vanishing condition fix $H_1 = 0$, so the zero momentum mode up to $\o$ order
becomes
\be
\ps = d_1 \ f^{-  i \fr{\o}{4 r_h^2} } ,
\ee
where $H_0 = d_1$ is an integration constant which will be determined by other boundary
condition.
The nonzero momentum mode $H_2$ is usually coupled to
$g^x_t$ and $a_x$, which can be determined from \eq{eq:constraint} to be
\be				\la{res:nearhorizongxy}
 H_2 =
c_1  \omega k \ f^{i \fr{\o}{4 r_h^2} }  \int_{r_h}^r d r \  \frac{ \ f^{-  i \fr{\o}{4 r_h^2} }  }{\sqrt{2} \ r \left(r^2+r_h^2\right)} + cc ,
\ee
where $cc$ is another integration constant. Due to the vanishing condition of $H_2$ at the horizon,
$cc$ should be zero.
Although the analytic form of $H_2$ can not be fixed, one can still find
a perturbative expansion form in the hydrodynamic limit which is sufficient to determine
the hydrodynamic coefficient (see below).

Now, let us investigate the asymptotic behavior of solutions.
Assuming the asymptotic forms of solutions as
\bea
a_x = A \ r^{\a} \quad {\rm and} \quad \Ph = B \ r^{\b} ,
\eea
\eq{eq:solve1} and \eq{eq:solve2} are reduced to
\bea
0 &=& \a (\a -2)  A \ r^{\a-2} + q B \ r^{\b+1} , \nn
0 &=& 2 q \a A \ r^{\a-5} + (\b^2 + 6 \b + 9 -q^2) B \ r^{\b-2} .
\eea
In order to have nontrivial solutions with non-vanishing $A$ and $B$, $\b$ should be
$\b = \a -3$ which determines $\a$ as
\be
\a = 1 \pm \sqrt{1 + q^2} ,
\ee
where $q^2 = 8$ for $z=2$ and $B$ is related to $A$
\be			\la{rel:asymptcoeff}
B = - \fr{\a (\a -2)}{q} A .
\ee
These results show that the asymptotic behaviors of $a_x$ and
$g^x_t$ ($= \int dr \ \Ph$) are described by
\bea			\la{sol:asympbeh}
a_x &=& a_0 \ r^4  \ls 1 + \cdots \rs + \fr{a_0^*}{r^2}  \ls 1 + \cdots \rs, \nn
g^x_t &=& g_{t0} \ r^2  \ls 1 + \cdots \rs + \fr{Bg_{t0}^*}{r^4}  \ls 1 + \cdots \rs  ,
\eea
where $a_0 = A$, $g_{t0}= B/2$ and ellipsis implies the higher order corrections.
From \eq{eq:zeromom}, the asymptotic form of the zero momentum mode in $g^x_y$
can be easily determined to be
\be
\ps = g_{y0}  \ls 1 + \cdots \rs + \fr{g_{y0}^* }{r^4}  \ls 1 + \cdots \rs .
\ee
The asymptotic behavior of the nonzero momentum mode can be fixed
by inserting the asymptotic solutions in \eq{sol:asympbeh} into the constraint equation \eq{eq:constraint}, which leads
\be
H_2 =  \td{C} - \fr{c_1 \o k }{2 \sqrt{2} r^2}   ,
\ee
where $\td{C}$ is another integration constant.
Since $H_2$ is the nonzero momentum mode, $\td{C}$ should be zero.
This result is consistent with \eq{res:nearhorizongxy}
in the asymptotic region.
In sum, the asymptotic behaviors of fluctuations are
\bea			\la{res:asymbeh}
a_x &\sim& a_0 \ r^4 ,  \nn
g^x_t &\sim& g_{t0} \ r^2 , \nn
g^x_y &\sim&  g_{y0} ,
\eea
where coefficients, $a_0$, $g_{t0}$ and $g_{y0}$, correspond to sources of each
fluctuations and should be determined by appropriate asymptotic boundary conditions.
Comparing the near horizon solutions in \eq{sol:nearhorizonsol1} and \eq{sol:nearhorizonsol2}
with the asymptotic behaviors of fluctuations in \eq{res:asymbeh},
these coefficients are related to undetermined parameters, $c_1$ and $d_1$,
\bea			\la{res:coeffbv}
a_0 &=& \frac{c_1}{32 \ r_h^2} \lb 32 r_h^2-8 i  \ls 2 \log 2 -1 \rs  \omega +\left(\pi ^2-2\right) k^2 \rb ,\nn
g_{t0} &=&  -\frac{c_1}{16 \sqrt{2} \ r_h^2} \lb 32 r_h^2-8 i  \ls 2 \log 2 -1 \rs \omega  +\left(\pi ^2-2\right) k^2 \rb,\nn
g_{y0} &=& d_1 .
\eea
These asymptotic behaviors of solutions are totally different from the AdS ones.
On the AdS background, as mentioned previously, $a_0$ is independent with $g_{t0}$
and $g_{y0}$ because of the decoupling of the vector fluctuation from the others.
The $z=2$ Lifshitz geometry has a different situation in which the asymptotic value of
$g^x_y$ is independent from others while $a_0$ is proportional to $g_{t0}$.
This fact, as will be shown, leads to totally different retarded Green functions.

In order to evaluate the retarded Green function, we first evaluate the boundary term
from the bulk action which reduces to
\be
S_B = \frac{1}{16 \pi G} \int d^{3} x
\ls  -  r^5 \  g_{y0}  \  {g^x_y}'  +  r^5 \ g_{t0} \
 {g^x_t}'  - r^3  a_0  \ a_x '   \rs .
\ee
From the boundary term together with \eq{res:coeffbv}, after discarding the contact 
terms,
the resulting retarded Green functions of the non-relativistic Lifshitz theory read
\bea
\bra J^x \ J^x \ket &=&
\frac{r_h^4 \lb 192 i \omega  r_h^2 + \left(16 r_h^2 +192  r_h^2 \log r_h -27 i \omega \right) k^2  \rb }{72 \pi  G \lb 32 r_h^2 -8 i \omega  (2 \log 2-1)+\left(\pi
   ^2-2\right) k^2 \rb}    , \nn
\bra T^t_x  \ T^t_x \ket &=&
\frac{ r_h^4 \lb 192 i \omega  r_h^2  - \left(80 r_h^2
- 192  r_h^2 \log  r_h   +
9 i \omega \right) k^2 \rb  }{72 \pi  G \lb 32 r_h^2 -8 i \omega  (2 \log 2-1)+\left(\pi
   ^2-2\right) k^2 \rb}   ,\nn
\bra T^y_x  \ T^y_x \ket &=& \frac{i \omega  r_h^2}{16 \pi  G}, \nn
\bra J^x  \ T^y_x \ket &=& \frac{\sqrt{2}  \omega k \ r_h^4}{\pi  G \lb 32 r_h^2 -8 i \omega  (2 \log 2-1)+\left(\pi
   ^2-2\right) k^2 \rb}  , \nn
\bra T^t_x  \ T^y_x \ket &=& - \frac{ \omega  k \ r_h^4}{\pi  G \lb 32 r_h^2 -8 i \omega  (2 \log 2-1)+\left(\pi
   ^2-2\right) k^2 \rb}  , \nn
\bra J^x  \ T^t_x \ket &=& 0 .
\eea
From the retarded Green function of the tensor mode $T^y_x$, the shear viscosity
is given by
\be
\et \equiv \lim_{\o ,k \to 0} \fr{\bra T^y_x  \ T^y_x \ket}{i \o} = \fr{r_h^2}{16 \pi G} ,
\ee
which shows the consistent result with the zero momentum calculation.
Using the fact that the entropy density $s$ is given by $r_h^2 / 4 G$ in \eq{res:thentropy},
the shear viscosity of the dual non-relativistic Lifshitz theory saturates the lower bound of
$\et/s \ge 1/4 \pi$ \cite{Iqbal:2008by}.  As shown in \cite{Park:2013goa}, the DC conductivity leads to
\be
\s \equiv \lim_{\o ,k \to 0} \fr{\bra J^x \ J^x \ket}{i \o} = \fr{r_h^4}{12 \pi G } .
\ee
In terms of Hawking temperature, these results can be rewritten as
\bea
\et &= &  \fr{1}{16 G}  T_H , \nn
\s &= & \fr{\pi }{12 G }  T_H^2 ,
\eea
which shows that the shear viscosity and DC conductivity increases with
Hawking temperature linearly and quadratically.

Interestingly, the above retarded Green functions indicate the existence of a massive
quasi-normal mode whose dispersion relation is deviated from an usual Drude formula
$i \o = {\cal D} k^2$. First, the current-current
and momentum-momentum correlator have the same pole structure. Secondly,
this quasi-normal mode has the following dispersion relation
\be
i \o = \fr{4 r_h^2}{ 2 \log 2 -1 } + \fr{\pi^2 -2 }{8 (2 \log 2 -1 )} \ k^2 .
\ee
If one ignore the temperature-dependent part ($T_H \sim r_h^2$), it reduces to an
usual Drude formula. 
In this case, the momentum diffusion constant is given by
a constant which is the expected result due to the same scaling behavior
of $\o$ and $k^2$ \cite{Park:2013goa}. The ignored term is a new one which does not
appear in the holographic dual of the relativistic (conformal) field theory.
For understanding this result further,
we rewrite the above dispersion relation as a Shr\"{o}dinger-type equation
\be			\la{eq:shrodinger}
i \ \ls i \fr{\pa}{\pa t} \rs \ \ps (t,y)   =  - {\cal D} \ \fr{\pa^2}{\pa y^2} \ \ps (t,y)  + m \ \ps (t,y)  ,
\ee
with
\bea
{\cal D} &=& \fr{\pi^2 -2 }{8 (2 \log 2 -1 )}  , \nn
m &=& \fr{4 \pi T_H}{ 2 \log 2 -1 } ,
\eea
where an additional imaginary number on the left hand side denotes  
the instability
of the wave function. Here we can identify $m$ with an effective thermal mass,
because it linearly depends on Hawking temperature, and ${\cal D}$ with a
charge or momentum diffusion constant which is independent of Hawking temperature.
Solving the Shr\"{o}dinger equation, the wave function describing a quasi-normal mode at a given
temperature reads
\be
\ps (t,y) \sim e^{- \ls m + {\cal D} k^2 \rs t }e^{i k y} ,
\ee
which allows to reinterpret a quasi-normal mode
as a massive decaying mode.
This unstable mode dissipates more rapidly with increasing temperature and momentum.
Note that the hydrodynamic calculation is valid only in the range of $\o \sim k^2 \ll T_H$.
Another interesting thing in the non-relativistic Lifshitz medium is that there exists a quasi-normal mode
even at zero temperature which follows the usual Drude formula.


\section{Discussion}

We have studied the holographic renormalization of the Einstein-Maxwell-dilaton
theory, which is the dual of the Lifshitz field theory, and its hydrodynamics. For the finiteness of the
boundary action and energy-momentum tensor, 
we have introduced appropriate local counter terms. There are
infinitely many possibilities which make the on-shell gravity action finite.
In this paper, we started with five lowest order counter terms. After imposing the non-vanishing value
to $c_0$, which is the counter term for the AdS geometry, we investigated the
holographic renormalization of the dual Lifshitz field theory. In this case, due to the redundancies
of the constraints five coefficients are not fully determined. Although we can
get rid of such redundancies by setting two of them to be zero by hand, there 
still remains an ambiguity in choosing them. In our set-up, six cases are possible and
all cases, when the boundary is located at infinity, give rise to the same thermodynamics
consistent with the Lifshitz black brane thermodynamics. If we further take into account the
holographic renormalization flow of this system, since IR physics is usually depend on the
local counter terms, the IR theory can not be uniquely determined without additional constraints. 
Those additional constraints would be related to the details of the dual Lifshitz theory 
or the renormalization scheme dependence of the field theory. So it would be interesting
to study further the microscopic aspects of the dual Lifshitz field theory and its renormalization. 

We also investigated the hydrodynamics of the non-relativistic Lifshitz theory ($z=2$).
In the nonzero momentum limit, the transverse mode of the gauge field couples to the
shear mode of the metric fluctuation due to the nonzero background gauge field. 
This fact leads to nontrivial retarded Green functions between the current and momentum operator.
In the non-relativistic medium, interestingly 
the current-current, current-momentum and momentum-momentum correlators
show a massive quasi-normal mode unlike the relativistic cases, which is slightly deviated from the usual
Drude formula.
Its mass is linearly proportional to temperature and the momentum diffusion constant is independent
of temperature. Moreover, this quarsi-normal mode still remains even at zero temperature 
with the usual Drude formula, $i \o = {\cal D} k^2$.
It would be interesting to study whether the other Lifshitz models also
generate a similar massive quasi-normal mode.

\vspace{1cm}

{\bf Acknowledgement}

This work has been supported by the WCU grant no. R32-10130 and the Research fund no. 1-2008-2935-001-2
by Ewha Womans University.  
C. Park was also supported by Basic Science Research Program through the National Research Foundation of 
Korea(NRF) funded by the Ministry of Education (NRF-2013R1A1A2A10057490).


\end{document}